\documentclass[%
aip,
apl,
reprint,
groupedaddress]{revtex4-1}

\usepackage{helvet}
\usepackage{amsmath}
\usepackage{amssymb}
\usepackage{amsfonts}
\usepackage{graphicx}

\begin{document}

\title[]{Migration of Mg and other interstitial metal dopants in GaN}

\author{Giacomo Miceli}
  \email{giacomo.miceli@epfl.ch}
  \affiliation{Chaire de Simulation \`a l'Echelle Atomique (CSEA), %
  Ecole Polytechnique F\'ed\'erale de Lausanne (EPFL), %
  CH-1015 Lausanne, Switzerland}

\author{Alfredo Pasquarello}
  \affiliation{Chaire de Simulation \`a l'Echelle Atomique (CSEA), %
  Ecole Polytechnique F\'ed\'erale de Lausanne (EPFL), %
  CH-1015 Lausanne, Switzerland}

%
%
\begin{abstract}
The minimum energy paths for the migration of interstitial Mg in wurtzite GaN are
studied through density functional calculations. The study also comprises Li, Na,
and Be dopants to examine the dependence on size and charge of the dopant species.
In all cases considered, the impurities diffuse like ions without any tendency of
localizing charge. Li, Mg, and to some extent Na, diffuse almost isotropically in
GaN, with average diffusion barriers of 1.1, 2.1, and 2.5 eV, respectively. 
Instead Be shows a marked anisotropy with energy barriers of 0.76 and 1.88 eV for 
diffusion paths perpendicular and parallel to the $c$-axis. The diffusion barrier 
generally increases with ionic charge and ionic radius, but their interplay is not 
trivial. The calculated migration barrier for Mg is consistent with the values 
estimated in a recent $\beta^-$ emission channeling experiment.
\end{abstract}


\maketitle

%
%
Its wide and direct band gap, high thermal and electric conductivity, and large 
breakdown fields make of GaN an ideally suited compound for electronic and 
optoelectronic devices.\cite{pearton_JAP1999} However, while GaN is already an 
essential compound in commercially available blue-light-emitting diodes, higher 
concentrations of free carriers in both $n$- and $p$-type layers are required for a 
broader use of this material in electronic devices. While high electron densities
are routinely achieved through silicon doping,\cite{gotz_MSEB1999,sheu_JPCM2002}
the efficiency of $p$-type doping still lags behind and currently constitutes 
the major obstacle for further progress.

Magnesium substitutional to gallium has been hitherto recognized as the only 
effective $p$-type doping in GaN.\cite{amano_JJAP1989,nakamura_JJAP1992,nakamura_APL1994} 
However, the occurrence of self-compensation upon heavy Mg doping prevents one to
reach the required levels of hole densities.\cite{kaufmann_PRB2000,brochen_APL2013}
The precise origin of the self-compensation is still debated, but is likely 
associated to point defects.\cite{kaufmann_APL1998,kozodoy_JCG1998,%
vandewalle_JCG1998, hautak_PRL2003,latham_PRB2003,miceli_MEE2015,miceli_PRB2016}
A recent theoretical study has suggested the Mg interstitial (Mg$_\text{i}$)
to play a key role in this process.\cite{miceli_PRB2016,miceli_MEE2015} This 
proposal has subsequently received support from a $\beta^-$ emission channeling 
experiment, which provided a direct proof of the occurrence of Mg$_\text{i}$.%
\cite{wahl_PRL2017} To complete this picture, it is important to understand the 
diffusion properties of the Mg interstitial, which determine the device processing
procedures and the electrical properties of the grown samples.\cite{fahey_RMP1989}
The description of the Mg diffusion process in GaN achieved so far is highly 
inconsistent. Experimental investigations lead to a large spread of diffusion 
barriers ranging from 1.3 to 5 eV.\cite{chang_APL1999,pan_SSE1999,benzarti_JCG2008,%
kohler_JAP2013,wahl_PRL2017} Recent experimental estimates situate the transition
barrier in a fairly large interval ranging from 1.3 to 2.0 eV.\cite{wahl_PRL2017}
In addition, a theoretical study based on classical force fields yields activation
barriers lower than 0.7 eV, and is thus not helpful in sorting out the experimental
data.\cite{harafuji_PSSC2003}

%
%
In this Letter, we investigate the minimum energy paths and the transition barriers 
for the diffusion of the interstitial Mg impurity in wurtzite GaN using
density-functional calculations. For comparison, we also include in our study the
diffusion of Li, Na, and Be ions. Larger ionic radii or larger ionic charges 
generally lead to higher energy barriers. The diffusion is generally quite 
isotropic, except for Be$^{2+}$, which diffuses with particularly low barriers 
in directions perpendicular to the $c$-axis. The average energy barrier calculated
for Mg$^{2+}$ is 2.1 eV, in agreement with the range of values estimated 
in a recent experimental study.\cite{wahl_PRL2017}

%
%
In this work, the atomic geometries and the energetics of the impurities in GaN 
are determined within the framework of density functional theory based on the
generalized gradient approximation proposed by Perdew, Becke, and Ernzerhof (PBE).%
\cite{pbe} Our computational scheme relies on norm-conserving pseudopotentials 
and plane-wave basis sets as implemented in the {\sc q}uantum-{\sc espresso} 
software package.\cite{quantum_espresso} The kinetic energy cut-off for the 
plane-wave basis sets is set at 45 Ry. The cation interstitial impurities are 
modeled in 96-atom supercells of GaN. We use lattice parameters fixed 
at their experimental values ($a=3.189$ \AA\ and $c=5.185$ \AA, 
Ref.\ \onlinecite{madelung2004}), as they differ by less than 0.3\%\ from
the equilibrium PBE values. The Brillouin zone of the supercell is sampled
with one special {\bf k}-point lying off the $\Gamma$ point. The minimum energy
paths of cation diffusion are identified through the nudged-elastic-band (NEB) 
scheme.\cite{henkelman1_JCP2000} We adopted a climbing image to determine the 
geometries and the energy barriers at the transition states.\cite{henkelman2_JCP2000} 
A minimization algorithm is applied until the residual total forces acting on each
image in the direction perpendicular to the path are smaller than 0.05 eV/\AA. 
To test the convergence of our calculations, we also evaluate the activation 
energies for Be diffusion using a denser $2\times2\times2$ {\bf k}-point grid. 
Similarly, we examined the effect of $3d$ electrons included in the Ga valence 
shell. The activation energies are found to remain unchanged within 0.1 eV.
To estimate the effect of using experimental rather than theoretical lattice 
parameters, we focus on the energy difference between the impurity in the 
octahedral and in the tetrahedral site, and find equivalent values within 0.05 eV. 
Furthermore, we use the same energy difference to examine the long-range 
relaxation effects resulting from the use of a finite supercell in the case 
of Mg. Using a larger supercell of 289 atoms, we find agreement within 0.03 eV.

\begin{figure}
 \centering
 \includegraphics[width=8.6cm]{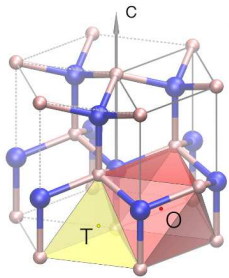}
 \caption{Crystal structure of wurtzite GaN with Ga atoms in pink and N atoms in
          blue. The Ga-based octahedral and tetrahedral interstitial volumes are
          highlighted in red and yellow, respectively. Points indicated by O and
          T correspond to their respective centers.}
 \label{fig:wurtzite}
\end{figure}

\begin{figure*}
 \begin{minipage}{1.25\columnwidth}
 \centering
 \includegraphics[width=12.9cm]{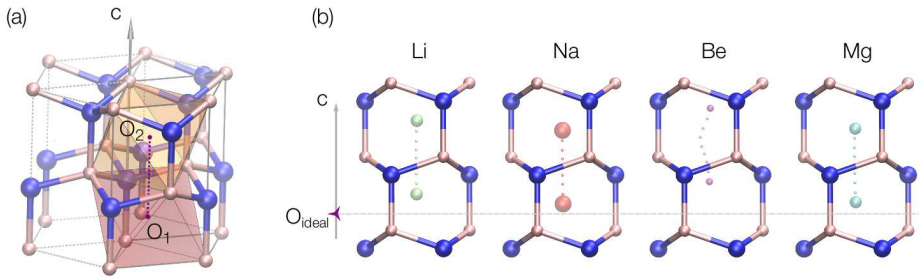}
 \end{minipage}
 \hfill
 \begin{minipage}{0.55\columnwidth}
  \caption{(a) Schematic representation of the diffusion path parallel to the 
           $c$-axis between two adjacent octahedral volumes. (b) Lateral view of
           the minimum energy diffusion path for Li$^+$, Na$^+$, Be$^{2+}$, and 
           Mg$^{2+}$, respectively.}
 \label{fig:pathc}
 \end{minipage}
\end{figure*}

\begin{table}
 \caption{
          Ionic radii ($r_{\rm ion}$) of Li$^{+}$, Na$^{+}$, Be$^{2+}$, and 
          Mg$^{2+}$ and structural parameters in their ground-state octahedral 
          site (O$^\prime$). Distances from nearest neighbor Ga and N atoms are 
          given. OO$^\prime$ gives the displacement along the $c$ direction with
          respect to the ideal octahedral site O shown in Fig.\ \ref{fig:wurtzite}.  
          All lengths are in \AA.}
 \label{tab:octa}
  \begin{ruledtabular}
  \begin{tabular}{lcccc}
   Cation    & $r_{\rm ion}$ & Ga      & N       & OO$^\prime$ \\
   \hline
   Li$^{+}$  & 0.76          & 2.34    & 1.91    & 0.62 \\
   Na$^{+}$  & 1.02          & 2.27    & 2.13    & 0.28 \\
   Be$^{2+}$ & 0.45          & 2.60    & 1.67    & 1.06 \\
   Mg$^{2+}$ & 0.72          & 2.35    & 2.03    & 0.35 \\
  \end{tabular}
 \end{ruledtabular}
\end{table}

%
%
The wurtzite structure achieves a tetrahedral-octahedral honeycomb space-filling
with either Ga or N atoms at the vertices of the polyhedra. As a matter of 
convenience, we illustrate in Fig.\ \ref{fig:wurtzite} the Ga-based tessellation. 
To determine the ground-state for interstitial impurities, we place the cations
at the centers of the interstitial polyhedra and allow for atomic relaxation 
until a locally stable structure is achieved. In all cases, the ground state 
is found for the cation in the position O$^\prime$ within the octahedral volume. 
The energy of the metastable state lying within the tetrahedral volume lies 
higher in energy by 1.10, 2.86, 0.57, and 2.04 eV for Li$^+$, Na$^+$, Be$^{2+}$, 
and Mg$^{2+}$, respectively. 

%
%
Within the octahedral volume, the ground-state site O$^\prime$ lies on the axis
of the hexagonal channel and thus preserves the axial symmetry of the wurtzite 
structure. For the investigated impurities, we give in Table \ref{tab:octa} the 
distances between the O$^\prime$ site and the nearest neighbor atoms of the GaN 
lattice, as well as its displacement OO$^\prime$ with respect to the ideal O site.
We observe that the Na$^+$ ion, which features the largest ionic radius (cf.\ 
Table \ref{tab:octa}), lies closest to the ideal O site, whereas the Be$^{2+}$ 
and Li$^+$, which have smaller ionic radii lie closer to the plane of the N anions.
A graphical view of the location of the O$^\prime$ site with respect to the atomic
planes is displayed in Fig.\ \ref{fig:pathc}(b). In particular, we obtain for 
Mg$^{2+}$ an OO$^\prime$ displacement of 0.35 \AA, to be compared with the shift
of $0.60\pm0.14$ \AA\ measured in Ref.\ \onlinecite{wahl_PRL2017}.

%
%
In the metastable tetrahedral site, the  Li$^+$, Na$^+$, and Mg$^{2+}$ cations 
are fourfold coordinated by nearest-neighbor N atoms and are aligned with the 
Ga and N lattice atoms in a column parallel to the $c$-axis. At variance, the 
Be$^{2+}$ ion finds its metastable position at the centers of the tetrahedron 
faces, where it can optimize its interactions with three nearest neighbor N 
atoms due to its small size.

%
%
The diffusion in GaN can be described by determining the minimum energy paths
between nearby octahedral sites. By comparing the formation energies
of charged and neutral species, we verified that the ionic state is always 
preserved along all the considered diffusion paths. Hence, there is no tendency 
to generate localized electronic states during the diffusion of the 
interstitial impurities.  

\begin{figure}
 \centering
 \includegraphics[width=8.6cm]{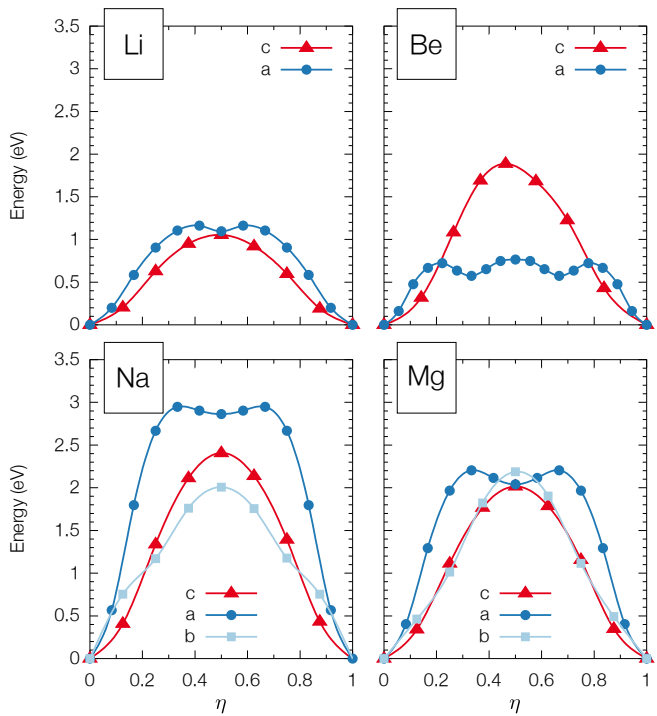}
 \caption{Minimum energy paths of migration for Li$^{+}$, Na$^{+}$, Be$^{2+}$,
          and Mg$^{2+}$ between octahedral sites along directions parallel 
          (path $c$) and perpendicular (paths $a$ and $b$) to the $c$-axis 
          of the wurtzite structure. The adimensional variable $\eta$ varies
          from 0 to 1 along the diffusion path.}
 \label{fig:energies}
\end{figure}

%
%
The interstitial ionic species can migrate through the open hexagonal channel 
parallel to the $c$-axis, which we denote as path $c$. As shown in Fig.\ \ref{fig:%
pathc}(a), the ionic species directly hop between adjacent octahedral sites by 
crossing the double GaN-layer perpendicular to the [0001] direction. In the case 
of Mg$^{2+}$, this diffusion path runs straight along the axis of the channel as 
shown in Fig.\ \ref{fig:pathc}(b). The lack of centrosymmetry in the wurtzite 
structure yields a non-symmetric minimum-energy-path profile with an energy barrier
of 2.01 eV. At the transition state, the coordinate of the Mg$^{2+}$ ion along 
the $c$-axis closely corresponds to plane of Ga atoms, with a Ga-Mg distance of 
2.26 \AA. The Li$^{+}$ and Na$^{+}$ ions diffusing along the $c$-axis show the 
same behavior [Fig.\ \ref{fig:pathc}(b)]. For these atomic species, we find energy
barriers of 1.05 and 2.41 eV and distances to Ga atoms of 2.07 and 2.17 \AA, 
respectively. 
The calculated diffusion barrier of Li$^{+}$ along the $c$ axis is lower
by 0.5 eV than obtained in a previous study with the local density 
approximation and with a smaller unit cell.\cite{bernardini_PRB2000}
By comparing the Li$^{+}$ and Na$^{+}$ ions, which both carry the 
same charge, one remarks that the larger ionic radius of the latter causes a 
significant increase in the ionic barrier. To estimate the effect of the charge, 
one can compare the diffusion of Mg$^{2+}$ and Li$^{+}$ ions, which feature similar 
ionic radii. It is seen that the larger ionic charge of the Mg$^{2+}$ ion leads 
to a higher barrier. For the Be$^{2+}$ ion, we observe that the lowest-energy 
path does not run along the axis of the hexagonal channel [see Fig.\ \ref{fig:%
pathc}(b)]. At the transition state, the Be$^{2+}$ ion shows two N atoms at a 
distance of 1.85 \AA\ and a third one at 2.05 \AA, leading to an energy barrier 
of 1.88 eV (cf.\ Fig.\ \ref{fig:energies}). This path is made possible because 
of the small ionic radius of the Be$^{2+}$ ion. Our results for the diffusion 
path of Be$^{2+}$ qualitatively agree with a previous study within the local 
density approximation,\cite{vandewalle_PRB2001} but the energy barrier calculated 
in this work is found to be lower by 0.89 eV. We carefully checked the convergence 
against all the computational parameters ensuring convergence of our result within 
0.1 eV. The origin of the higher barrier in Ref.\ \onlinecite{vandewalle_PRB2001} 
should thus be ascribed to the use of a different energy functional.

\begin{figure}
 \centering
 \includegraphics[width=8.6cm]{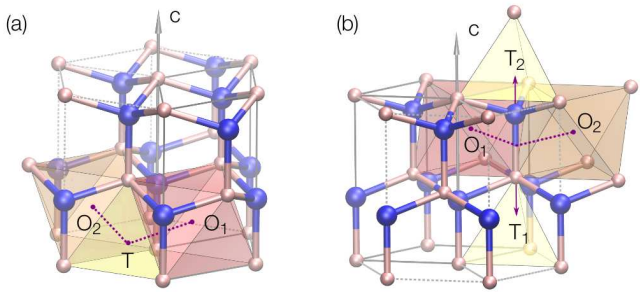}
 \caption{Diffusion paths in the plane perpendicular to the $c$-axis. (a) Diffusion
          path $a$: the interstitial impurity diffuses passing through a tetrahedral
          volume T. (b) Diffusion path $b$: concerted migration mechanism involving
          the breaking of a Ga-N bond of the lattice, in which the Ga and N atoms
          move into the tetrahedral volumes T$_1$ and T$_2$, respectively.} 
 \label{fig:pathab} 
\end{figure}

%
%
The ionic impurities can also diffuse in directions orthogonal to the $c$-axis.
The impurity can diffuse between two nearby octahedral volumes O$_1$ and O$_2$
passing through the tetrahedral volume T that connects them. This diffusion 
channel is denoted as path $a$ and is schematically illustrated in Fig.\ 
\ref{fig:pathab}(a). By symmetry, the paths O$_1$-T and T-O$_2$ are equivalent. 
For all diffusing species considered here, the minimum energy path passes through
the metastable tetrahedral site. For Mg$^{2+}$, Li$^+$, and Na$^+$, this leads to
a single transition state. We find respective energy barriers of 2.20, 1.16, and
2.95 eV (Fig.\ \ref{fig:energies} and Table \ref{tab:barriers}). The trends with 
ionic size and charge are the same as for the diffusion along path $c$. In the 
case of Be$^{2+}$, the metastable position in the tetrahedral volume lies off the
axis of the tetrahedron and three transition states occur upon the O$_1$-T-O$_2$
migration with very similar energy barriers of 0.72, 0.76, and 0.72 eV (Fig.\ 
\ref{fig:energies}). We assign this different behavior to the small size of 
Be$^{2+}$, leading to distances of only $\sim$1.61 \AA\ to the nearest N atoms 
at the transition states. In Ref.\ \onlinecite{vandewalle_PRB2001}, the energy 
barrier for Be$^{2+}$ along this path was found to be 1.18 eV, larger by 0.42 eV 
than the present finding, but not as different as found for path $c$.

%
%
We also identified a second nonequivalent diffusion channel for migration 
perpendicular to the $c$-axis, which we denote as path $b$. Unlike paths $c$ and 
$a$, this channel does not correspond to a sequence of jumps between interstitial
volumes, but implies a concerted mechanism, which involves the breaking of a Ga--N
bond of the lattice. As schematically illustrated in Fig.\ \ref{fig:pathab}(b), 
the interstitial impurity in the octahedral site heads straight onto the center 
of a Ga--N bond, causing the Ga and N atoms to move apart along the $c$ direction.
This movement is facilitated by the occurrence of the interstitial tetrahedral 
volumes T$_1$ and T$_2$, which can accommodate these atoms. At the transition 
state, the interstitial impurity, the Ga atom, and the N atom are vertically 
aligned along the $c$ direction [cf.\ Fig.\ \ref{fig:pathab}(b)]. After the 
transition state, the Ga--N bond is formed back and the diffusing impurity moves
to its ground state in the nearest octahedral volume. Path $b$ is found as a 
stable diffusion channel only for Mg$^{2+}$ and Na$^{+}$. The respective 
calculated energy barriers are 2.19 and 2.01 eV (see Table \ref{tab:barriers}).
In the plane perpendicular to the $c$-axis, the diffusion of Mg$^{2+}$ along path
$b$ shows approximately the same barrier as along path $a$. However, in the case
of Na$^{+}$, the energy barrier of path $b$ is lower than that of path $a$ by 
almost 1 eV. In the case of Be$^{2+}$ and Li$^{+}$, nudged-elastic-band calculations
started from path $b$ revert spontaneously to path $a$. These results indicate 
that path $b$ becomes viable only for impurities with either a large ionic radius
or a large ionic charge.

\begin{table}
 \caption{Energy barriers (in eV) for the migration of Li$^{+}$, Na$^{+}$, 
          Be$^{2+}$, and Mg$^{2+}$ along three different diffusion paths.}
 \label{tab:barriers}
  \begin{ruledtabular}
  \begin{tabular}{lccc}
   Cation    &  Path $c$ & Path $a$ & Path $b$ \\
   \hline
   Li$^{+}$  &  1.05     &  1.16    &   --   \\
   Na$^{+}$  &  2.41     &  2.95    &  2.01  \\
   Be$^{2+}$ &  1.88     &  0.76    &   --   \\
   Mg$^{2+}$ &  2.01     &  2.20    &  2.19  \\
  \end{tabular}
 \end{ruledtabular}
\end{table}

%
%
All the calculated energy barriers are collected in Table \ref{tab:barriers}. We
remark that Li$^{+}$ and Mg$^{2+}$ show almost the same energy barriers along 
paths parallel and perpendicular to the $c$-axis, resulting in close to isotropic
diffusion. This is true to a lesser extent for Na$^{+}$, for which the energy 
barriers differ up to $\sim$20\% from their average. The anisotropy is more 
pronounced in the case of Be$^{2+}$, for which the relative difference reaches 
42\%\ with respect to the average. For understanding the specific behavior of 
Be$^{2+}$, we draw a comparison with Mg$^{2+}$ along paths $c$ and $a$. Along 
path $c$, the impurity crosses sequentially a triangle of N atoms and one of Ga 
atoms, the latter being responsible for the energy barrier. Along path $b$, the 
impurity also crosses triangles of N and Ga atoms, but simultaneously. This should
lead to a lower energy barrier in the latter case, due to a more effective screening
of the N atoms at the transition state. Indeed, this explains the anisotropy found
for Be$^{2+}$.\cite{vandewalle_PRB2001} However, we do not see a similar reduction
of the energy barrier along path $a$ for Mg$^{2+}$, despite this ion carries the
same charge. Inspection of the transition state reveals that along path $a$ the 
transition of Mg$^{2+}$ requires the outward displacement of the N atoms, unlike
for the smaller Be$^{2+}$. This effect entails an energy cost, which opposes the
more effective Coulombic screening and leads to similar energy barriers for path
$c$ and $a$ in the case of Mg$^{2+}$. This comparison clearly emphasizes the 
intricate interplay between size and charge effects in determining the transition
barriers of such ionic species in GaN.

%
%
In conclusion, we studied the diffusion of Mg$^{2+}$ and other interstitial
cations in GaN using density functional calculations. We identified three 
nonequivalent diffusion channels: one parallel and two perpendicular to the 
$c$-axis of the wurtzite crystal structure. The energy barriers generally increase
with ionic radius and ionic charge, but their interplay leads to nontrivial 
effects. The energy barriers of Mg$^{2+}$ calculated in this work support 
experimental estimates of about 2 eV.\cite{benzarti_JCG2008,wahl_PRL2017}

Financial support is acknowledged from the Swiss National Science Foundation 
(Grant No.\ 200020-152799). We used computational resources of CSCS and CSEA-EPFL.

\end{document}